\documentclass[preprint]{revtex4}
\usepackage{graphicx,float}
\usepackage{amssymb,amsmath}

\begin{document}
\title{Double dark-matter admixed neutron star}

\author{Zeinab Rezaei\footnote{E-mail: zrezaei@shirazu.ac.ir}}

\affiliation{Department of Physics, Shiraz University, Shiraz 71454, Iran\\
Biruni Observatory, Shiraz University, Shiraz 71454, Iran}

\begin{abstract}
The dark matter in neutron stars can exist
from the lifetime of the progenitor or captured by this compact object.
The properties of dark matter entered the neutron stars through each step
could be different from each other.
Here, we investigate the structure of
neutron stars which are influenced by the dark matter in two processes.
Applying a generalization of two-fluid
formalism to three-fluid one and the equation of state from the rotational curves of galaxies,
we explore the structure of double dark-matter admixed neutron stars.
The behavior of the neutron and dark matter portions for these stars is considered.
In addition, the influence of the dark matter equations of state on the stars
with different contributions of visible and dark matter is studied.
The gravitational redshift of these
stars in different cases of dark matter equations of state is investigated.

\end{abstract}
\maketitle
\section{Introduction}

Dark matter (DM) in protostellar halos significantly changes the
current theoretical framework for the formation of the first stars \cite{Spolyar8,Casanellas9}.
In neutralino DM annihilation, the heat which overwhelms the cooling mechanism
impedes the star formation process and leads to dark star,
a giant hydrogen-helium star larger than $\sim 1 AU$
which is powered by DM annihilation instead of nuclear fusion \cite{Spolyar8,Spolyar9}.
Including an energy term based on DM self-annihilation during the
cooling and collapse of the metal-free gas, in halos hosting the formation
of the first stars, it has been shown that the feedback induced on the
chemistry of the cloud modifies the properties of the gas throughout the collapse \cite{Ripamonti}.
It has been argued that the first stars form inside DM halos with the mass $\sim 10^6 M_{\odot}$
whose initial density profiles are laid down by gravitational
collapse in hierarchical structure-formation scenarios \cite{Freese}.
Regardless of the nature of the DM and its interactions, during the formation of the
first stars, the baryonic infall compresses the DM \cite{Freese}.
The evolution of young low mass stars in their Hayashi track in the pre-main-sequence phase has been
explored \cite{Casanellas9}. It has been clarified that for high DM densities, these objects stop their
gravitational collapse before reaching the main sequence forming an equilibrium state
with lower effective temperatures. However, for the lower DM densities,
these protostars continue to collapse and forming a main sequence star burning hydrogen
but at a lower rate. Their calculations confirm that the effective temperature and
luminosity of these stars depend on the properties of DM particles.
The DM particles of the
parent halo are accreted in the proto-stellar interior by adiabatic contraction
and scattering/capture processes and then its
rate increases to a value that energy deposition from annihilations significantly
alters the pre-main-sequence evolution of the star \cite{Iocco}.
The rate of energy injection
by weakly interacting massive particles (WIMPs) determines the stellar structure and main sequence evolution \cite{Scott}. It has been found that if the energy injected from the annihilation of WIMPs
is the order of the nuclear burning, the capture of weakly interacting DM affects the
structure and evolution of stars on the main sequence.
The accretion of more value of DM onto the star
leads to the condition that DM heating becomes the dominant power source \cite{Spolyar9}. Different effects such as variety of particle
masses and accretion rates, nuclear burning, feedback mechanisms, and possible repopulation of DM
density due to capture on the stellar evolution have been considered \cite{Spolyar9}. The results indicate that the first
stars are very large, puffy, bright, and cool during the accretion.
Besides, as the DM fuel is exhausted, a heavy main sequence star forms which finally collapses
to a massive black hole. The evolution of accreting
stars has been explored using several gas mass accretion rates derived from cosmological simulations \cite{Hirano}.
In the dark star phase, the star can expand by over a thousand solar radius with the surface temperature below $10^4\ K$.
It has been found that a dense DM halo around a cluster
changes the positions of the cluster' stars in the H-R diagram and leads to brighter
and hotter turn-off point giving the cluster a younger appearance \cite{Casanellas11}.
In addition, the DM density gradient inside the stellar cluster leads to a broader main sequence, turn-off and red giant branch
regions. The presence of relatively small amounts of DM modifies the critical phases of stellar evolution \cite{Sandin}.
It has been concluded that the
contraction of protostar to the main sequence could be significantly
delayed by WIMP annihilation heating \cite{Natarajan}.

A star orbiting close enough to an adiabatically grown supermassive
black hole may capture WIMPs which results in
a luminosity due to annihilation of captured WIMPs comparable to or even more than the luminosity
from thermonuclear burning \cite{Moskalenko}. These WIMP burners are mostly degenerate electron
cores with a very high surface temperature. The effect of weakly-interacting DM
accretion onto the stars and its annihilation on the main sequence stars
has been considered \cite{Fairbairn}.
The evolution of Population III stars with the capture and annihilation of WIMPs
has been explored \cite{Taoso8}. The results show that for enough dense DM, the annihilation
of WIMPs captured by Population III stars significantly changes the evolution of these stars.
It has been verified that the DM particles in the solar core affect the absolute frequency
values of gravity modes \cite{Lopes10}. The reduction of solar neutrinos flux due to the energy
dissipation by DM particles from the Sun and the consequent reduction of the
solar core temperature have been studied \cite{Taoso10}.

Because of the compactness and high density of compact objects, the accretion of DM particles can
take place on compact stars with more possibility than the normal stars \cite{Bertone,Kouvaris10}.
The accretion of dark matter
onto neutron stars and the effects of annihilating dark matter
on the temperature of these stars have been studied \cite{Lavallaz81}.
Assuming the DM
as a Fermi-gas with repulsive interaction between the DM particles which
scattered into the compact
star it has been shown that the smaller the
DM particle mass, the harder the EoS or the larger the maximum mass \cite{Li}.
The observational
consequences of the capture of DM onto astrophysical
compact objects have been considered \cite{Bertone}.
The role of a neutron
star progenitor on the accretion of WIMPs onto the
 neutron star has been investigated \cite{Kouvaris10}.
It has been concluded that the accretion
of a two-component millicharged dark matter onto a neutron
star can change the braking index of a pulsar \cite{Kouvaris14}.
The Self-annihilation of DM candidates in the central regions of
a typical neutron star has been proposed \cite{Perez-Garcia10}.
Some observational results related to extraordinary compact neutron stars are explained by the existence of dark matter
in these neutron stars \cite{Ciarcelluti695}.
Using a general relativistic two-fluid formalism and applying
non-self-annihilating DM particles, it has been found that
the structure of dark matter neutron stars depends on the DM
fluid and the size of the DM core \cite{Leung107301,Leung103528}.
The influence of mirror dark matter on the formation and structure of neutron stars
has been studied \cite{Sandin}.
Supposing that the fermionic DM and neutron matter have only gravitational interactions,
it has been confirmed that
the mass-radius relation depends on the mass of DM particles,
the amount of DM, and interactions between DM particles \cite{Xiang}.
The effects of bosonic DM on the structure
of compact stars have been explored \cite{Brito}.
Employing the relativistic mean-field theory in neutron stars with fermionic dark matter
which interacts directly with neutrons by exchanging standard model Higgs bosons,
 the equation of state has been computed \cite{Panotopoulos}.
The structure of non-rotating and rotating configurations of pure hadronic stars
with self-interacting fermionic asymmetric dark matter has been
 also calculated \cite{Mukhopadhyay}.

Regarding the above discussions, the DM can both exist from the first
steps of star formation or captured by the neutron stars from the environment. Therefore, the neutron star may be affected
by the DM in a double process (double dark-matter admixed neutron star). It should be noted that
the amount of DM captured by the star \cite{Sandin} and the mechanisms for accumulating \cite{Ciarcelluti695}
depend on the nature of DM. Therefore, in a double dark-matter admixed neutron star,
two kinds of DM with different equations of state (EOSs) can exist.
In this work, we are interested
in the properties of neutron stars which are double dark-matter admixed with
the same and also different DM EOSs.

\section{General relativistic three-fluid formalism}\label{drtff}

In order to study the structure of neutron stars which are admixed by the DM
through two different processes, we generalize the two-fluid formalism \cite{Sandin,Ciarcelluti695} to a three-fluid one for the neutron matter and two systems of DM.
In this model, a system composed of neutrons and two systems containing DM interact with each other just through gravity. We investigate the properties of a neutron star with three concentric
spheres. One of the spheres contains neutron matter, neutron sphere, and the other
two spheres are formed by DM, dark sphere number 1 (DS1) and dark sphere number 2 (DS2).
 Considering the static and spherically symmetric space-time with the line element, (together with the units in which G = c = 1),
\begin{eqnarray}
      d\tau^2=e^{2\nu(r)}dt^2-e^{2\lambda(r)}dr^2-r^2(d\theta^2+sin^2\theta d\phi^2),
 \end{eqnarray}
and the energy-momentum tensor of a perfect fluid,
\begin{eqnarray}
      T^{\mu \nu}=-p g^{\mu \nu}+(p+\varepsilon)u^{\mu}u^{\nu},
 \end{eqnarray}
we obtain the structure of the neutron star. In the above equation, $p$ and $\varepsilon$ denote the total
pressure and total energy density, respectively, which are related to the pressure and energy
density of three fluids,
\begin{eqnarray}\label{ps}
      p(r) = p_N(r) + p_{D1}(r)+ p_{D2}(r),
 \end{eqnarray}
   \begin{eqnarray}\label{eps}
      \varepsilon(r) =\varepsilon_N(r) + \varepsilon_{D1}(r)+ \varepsilon_{D2}(r).
 \end{eqnarray}
In Eqs. (\ref{ps}) and (\ref{eps}), $p_i$ and $\varepsilon_i$ present the pressure and energy density of neutron matter (i = N),
DM in DS1 (i = D1), and DM in DS2 (i = D2)
at position r from the center of the star, respectively. Using the above relations,
the Einstein field equations lead to \cite{Sandin,Ciarcelluti695,Xiang},
\begin{eqnarray}
      e^{-2\lambda(r)}=1-\frac{2M(r)}{r},
       \end{eqnarray}
\begin{eqnarray}
      \frac{d\nu}{dr}=\frac{M(r)+4\pi r^3 p(r)}{r[r-2M(r)]},
       \end{eqnarray}
\begin{eqnarray}
            \frac{dp_N}{dr}=-[p_N(r)+\varepsilon_N(r)] \frac{d\nu}{dr},
 \end{eqnarray}
\begin{eqnarray}
            \frac{dp_{D1}}{dr}=-[p_{D1}(r)+\varepsilon_{D1}(r)] \frac{d\nu}{dr},
 \end{eqnarray}
\begin{eqnarray}
            \frac{dp_{D2}}{dr}=-[p_{D2}(r)+\varepsilon_{D2}(r)] \frac{d\nu}{dr},
 \end{eqnarray}
in which $M(r)=\int_0^r dr 4 \pi r^2 \varepsilon(r)$ shows the total mass inside a sphere with radius $r$. The above relations are the result
of the assumption that the fluids interact just via gravity. These three-fluid TOV equations can be applied to calculate the neutron star structure and the properties of the neutron and dark spheres.
For the neutron matter with $\rho\gtrsim0.05 fm^{-3}$, we apply the neutron matter EOS sly230b in the Skyrme framework \cite{Chabanat}.
Moreover, for densities $\rho\lesssim0.05 fm^{-3}$, the EOS calculated by Baym \cite{Baym} is used. In addition, the EOSs of DM will be explained
in the following.
The boundary condition $p(R) = 0$
determines the radius, $R$, and the mass, $M(R)$, of the star. In addition, the radius and mass of each sector
are obtained with the conditions $p_N(R_N) = 0$, $p_{D1}(R_{D1})= 0$, and $p_{D2}(R_{D2})= 0$. It is important to note that the
pressure and density profiles of three fluids are different with each other.

\section{Dark mater equations of state}

To investigate the structure of double dark-matter admixed neutron stars, we should add
the DM EOSs to the general relativistic equations in section (\ref{drtff}). We are
interested in neutron stars which are affected by DM through two different processes. Therefore, we assume that the DM EOS of DS2 compared to the DM EOS of DS1 can be softer, stiffer, or the same.
  \begin{figure*}[t]
\centering{%
{
 \includegraphics[width=0.9\textwidth]{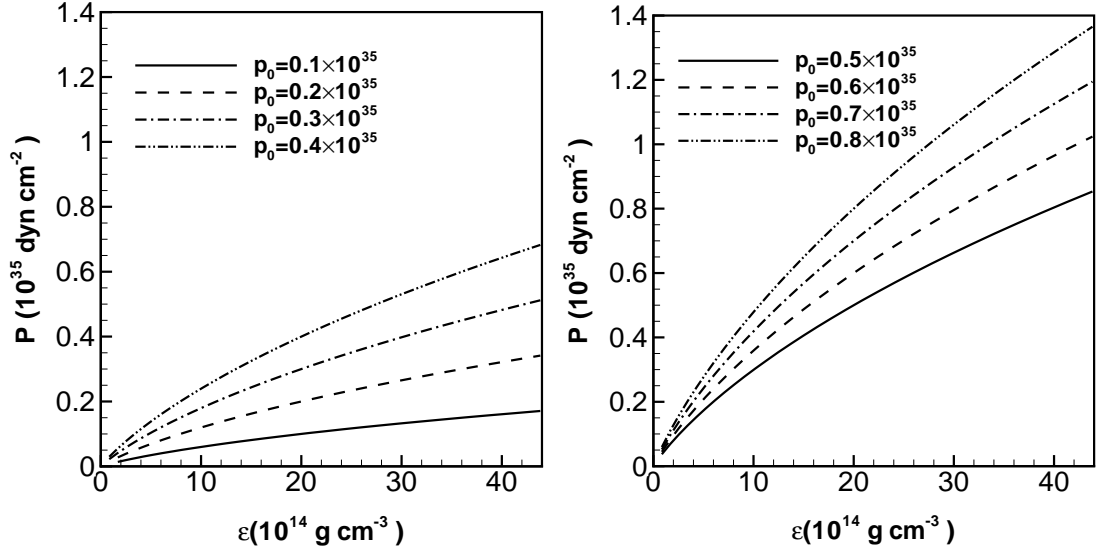}\qquad}}
\caption{Dark matter equation of state from the rotational curves of galaxies with
$\rho_0=0.2\times10^{16}g cm^{-3}$.}
\label{fig2}
\end{figure*}
To describe the DM in dark spheres, we employ the DM EOS
calculated using the observational data of the rotational curves of galaxies \cite{Barranco}.
The pseudo-isothermal model gives a mass density profile with the property of
regularity at the origin. Applying the velocity profile, geometric potentials, and gravitational potential,
the EOS resulted from the pseudo-isothermal density profile will have the form \cite{Barranco},
\begin{eqnarray}\label{213}
       {p}({\rho})=\frac{8  {p}_0}{\pi^2-8}[\frac{\pi^2}{8}-\frac{arctan\sqrt{\frac
       {{\rho}_0}{{\rho}}-1}}{\sqrt{\frac{{\rho}_0}
       {{\rho}}-1}}
	-\frac{1}{2}(arctan\sqrt{\frac
       {{\rho}_0}{{\rho}}-1}\ )^2],
 \end{eqnarray}
in which $\rho$ and $p$ represent the density and pressure of DM. The free parameters $\rho_0$ and $p_0$
correspond to the central density and pressure of galaxies. This EOS has a functional dependence
which is universal for all galaxies. It is possible to predict the central pressure and density of
the galaxies employing this EOS and the rotational curve data.
This universality for the galaxies could be extended to the other DM halos beyond DM halo of
galaxies. We mean that the DM halo of protostellar systems, normal stars, clusters, and
compact stars belong to this universality class. In such situation, the DM EOS of each halo is
 described by Eq. (\ref{213}) with the free parameters $\rho_0$ and $p_0$ that are equal to
the central density and pressure of that halo, respectively. Since in the present
study on neutron stars, we are dealing with the DM halo of neutron stars, we consider this universality
for these stars with the appropriate free parameters.
In addition, by comparing with the observational data, we have recently shown that
the DM in neutron stars can be described by this EOS \cite{Rezaei}. Therefore, to do this,
the values of $\rho_0$ and $p_0$ should be of the order of the central density and pressure of neutron
stars, i.e. $\sim10^{15}g cm^{-3}$ for the density and $\sim10^{34}dyncm^{-2}$ for the pressure.
Fig. \ref{fig2} shows the DM EOSs calculated using the rotational curves of galaxies
with the free parameters
$\rho_0=0.2\times10^{16}g cm^{-3}$ and different values of $p_0$, in the order of the
central density and pressure of neutron stars.
It is clear that the EOS with higher value of $p_0$ is stiffer.

\section{Structure of double dark-matter admixed neutron star}

\subsection{Total mass and visible radius}
\begin{figure*}[t]
\centering{%
{
   \includegraphics[width=0.9\textwidth]{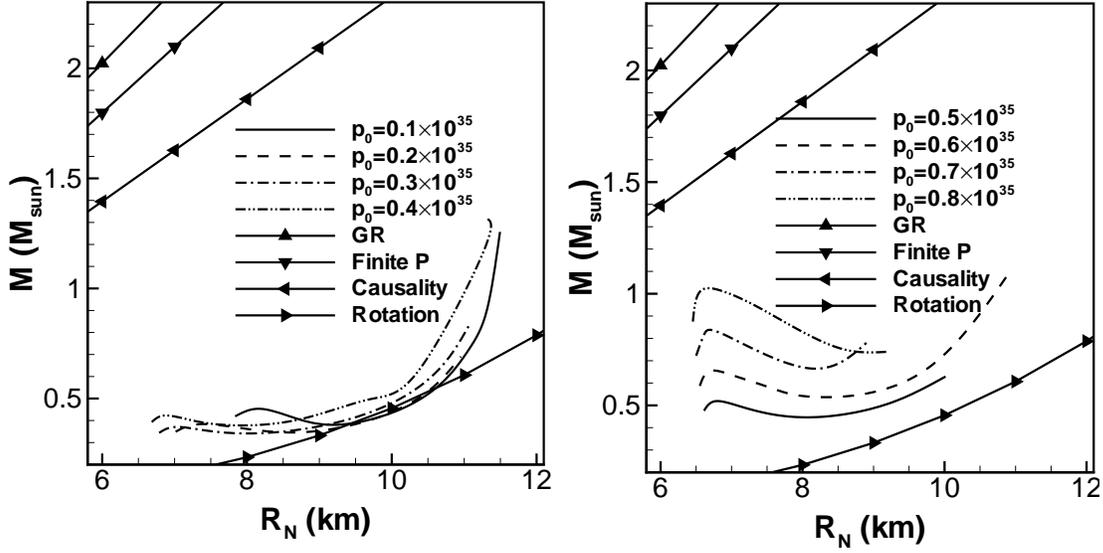}\qquad}}
\caption{Total mass of neutron star, $M$, versus the radius of neutron sphere, $R_N$, with the DM EOS $p_0=0.4\times10^{35}dyncm^{-2}$ for the DS1 and different DM EOSs for the DS2.
In addition, the curves which show the permitted regions are presented, see the text.}
\label{fig3}
\end{figure*}
Fig. \ref{fig3} presents the total mass of double dark-matter admixed neutron star versus
the visible radius, i.e. the radius of the neutron sphere, with the DM
EOS $p_0=0.4\times10^{35}dyncm^{-2}$ for the DS1 and different DM EOSs for the DS2.
We assume that the DM EOSs for the DS2 can be softer ($p_0<0.4\times10^{35}dyncm^{-2}$) than, the same ($p_0=0.4\times10^{35}dyncm^{-2}$), or stiffer ($p_0>0.4\times10^{35}dyncm^{-2}$) than the DM EOS of DS1. Therefore, we can study different dark-matter admixed neutron stars which are doubly affected by the softer, the same, or the stiffer DM compared to the primitive DM. We have also shown the permitted
region by presenting the curves from the general relativity $M>\frac{ R}{2 }$ (GR),
finite pressure $M>\frac{4R}{9}$ (Finite P), causality $M>\frac{10R}{29}$ (Causality),
and rotation of 716 Hz pulsar J1748-2446ad (Rotation) \cite{Lattimer}.
Our results confirm that almost all the stars with the above EOSs can exist.
The stars with the identical DM EOSs for two dark spheres are self-bound stars.
This behavior also stands for the softer DM EOSs. However, considering the DM in the DS2
be stiffer than the DM of DS1 leads to a mass-radius relation different from the previous cases. With the stiffer DM EOSs, the stars are gravitational-bound stars. We can conclude that
the behavior of the star depends on the kind of the dark matter which absorbs.
It is obvious that the stars with the stiffer DM in DS2 can be smaller and more massive, i.e. more compact. According to our results, the total mass and the radius of neutron sphere satisfy the Buchdahl condition,
i.e. $M < 4R_N/9$,  \cite{Schutz} and therefore its stability is confirmed.

\subsection{Total mass versus the radii of dark matter spheres}

\begin{figure*}[t]
\centering{%
{   \includegraphics[width=1.0\textwidth]{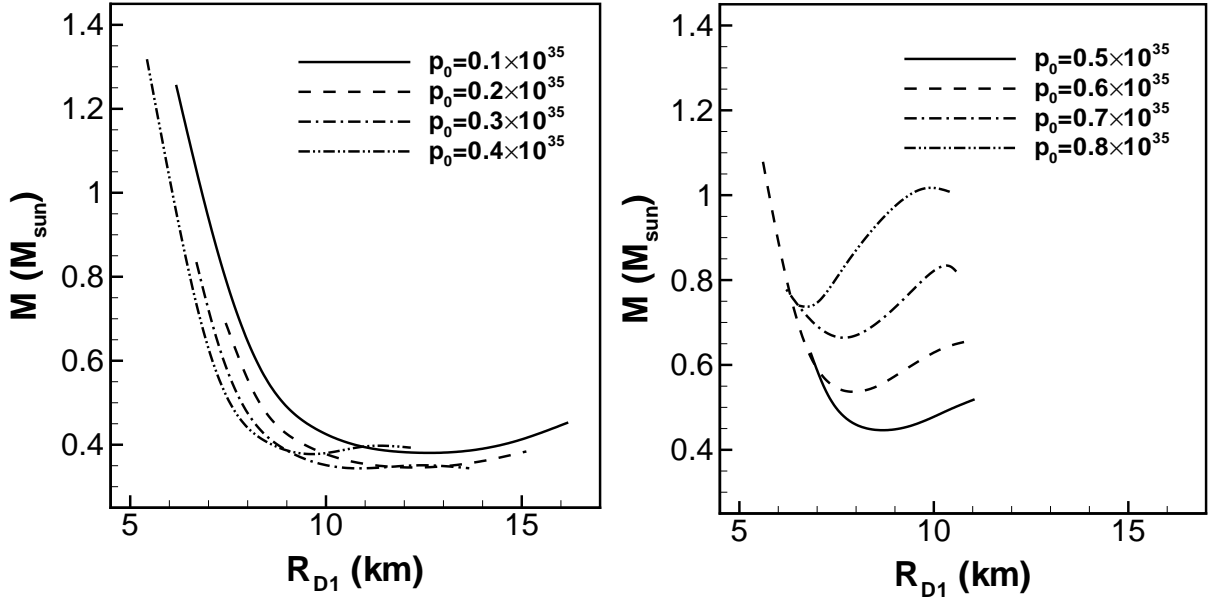}\qquad}}
\caption{Total mass of neutron star, $M$, versus the radius of DS1, $R_{D1}$, with the DM EOS $p_0=0.4\times10^{35}dyncm^{-2}$ for the DS1 and different DM EOSs for the DS2.}
\label{fig5}
\end{figure*}

Fig. \ref{fig5} presents the total mass of neutron star
versus the radius of DS1 with the DM
EOS $p_0=0.4\times10^{35}dyncm^{-2}$ for the DS1 and different DM EOSs for the DS2.
It can be seen that for the stars with identical DM in its dark spheres, the radius of
DS1 increases by decreasing the star mass. It is also true if the DM in DS2 is softer than
the DM in DS1 (Fig. \ref{fig5} Left). However, by increasing the stiffness of DM in DS2 compared to DM in DS1, the behavior of the star changes. In these cases (Fig. \ref{fig5} Right), the more massive stars have DS1 with larger radius. Fig. \ref{fig5} also confirms
that the stiffer DM in DS2 results in a smaller range for the radius of DS1. Moreover, the effect of the stiffness of DM in DS2 on the star mass depends on the DM in DS1. For DM EOSs with $p_0<0.4\times10^{35}dyncm^{-2}$, i.e. softer than the DM in DS1, the star mass decreases with the increase in the stiffness of DM in DS2 (Fig. \ref{fig5} Left). This is while for DM EOSs with $p_0>0.4\times10^{35}dyncm^{-2}$, i.e. stiffer than the DM in DS1, the star mass grows with increasing the stiffening of DM in DS2 (Fig. \ref{fig5} Right).
Figs. \ref{fig3} Left and \ref{fig5} Left show that for more massive stars, the
DS1 is surrounded by the neutron matter sphere. However, in the stars with lower masses,
the DS1 has a size larger than the neutron matter sphere. On the other hand, Figs. \ref{fig3} Right and \ref{fig5} Right confirm that for the DM EOSs with $p_0>0.4\times10^{35}dyncm^{-2}$, the condition is vice versa. In these cases, for more massive stars, the DS1 surrounds the neutron matter sphere and in the stars with lower masses the DS1 is hidden in the neutron matter sphere.
The total mass of neutron star and the radius of DS1 also provide the Buchdahl condition, i.e. $M < 4R_{D1}/9$,
and the stability is verified.

\begin{figure*}[t]
\centering{%
{
   \includegraphics[width=1.0\textwidth]{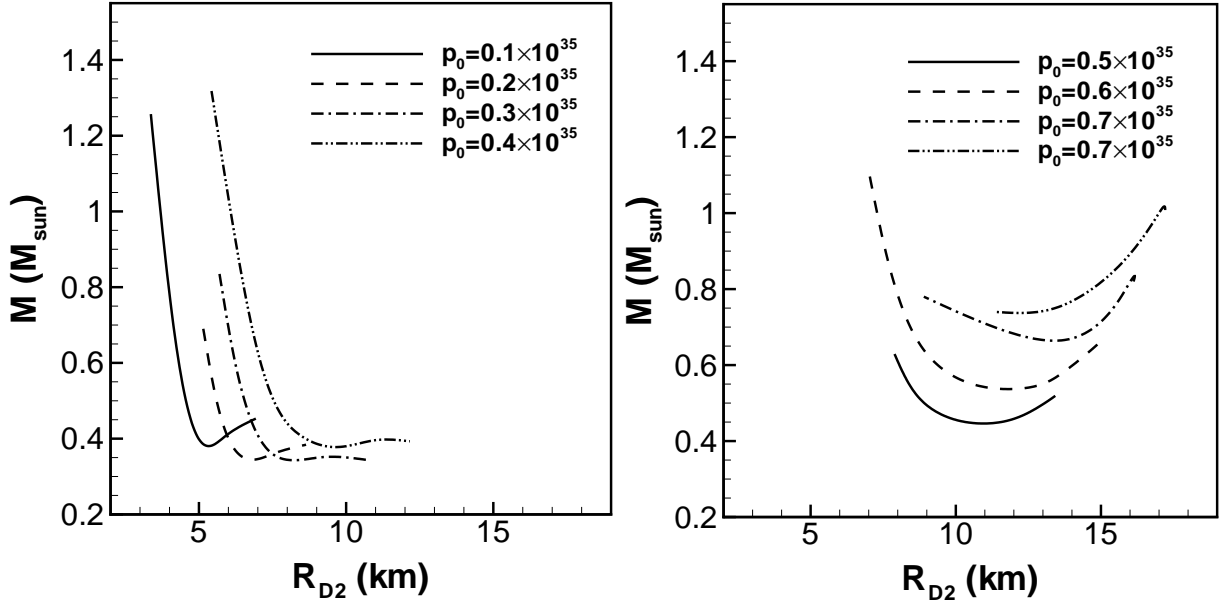}\qquad}}
\caption{Total mass of neutron star, $M$, versus the radius of DS2, $R_{D2}$, with the DM EOS $p_0=0.4\times10^{35}dyncm^{-2}$ for the DS1 and different DM EOSs for the DS2.}
\label{fig7}
\end{figure*}

Fig. \ref{fig7} shows the total mass of neutron star
as a function of the DS2 radius for the stars with the DM
EOS $p_0=0.4\times10^{35}dyncm^{-2}$ for the DS1 and different DM EOSs for the DS2.
Similar to the DS1, the size of DS2 is
larger for the stars with lower masses and softer DM EOS for the DS2.
It is clear that with the stiffer DM EOS for the DS2, the extension of DS2 is more considerable. The $M-R_{DS2}$ relation depends on the DM EOS stiffness of DS2. With the comparison of Figs. \ref{fig5} and \ref{fig7}, it can be found that for DM EOSs with $p_0<0.4\times10^{35}dyncm^{-2}$ the DS2 is within the DS1. However, for DM EOSs with $p_0>0.4\times10^{35}dyncm^{-2}$, the DS2 surrounds the DS1. It is clear from Figs. \ref{fig3} and \ref{fig7} that for massive stars with $p_0<0.4\times10^{35}dyncm^{-2}$, the DS2 is embedded within the neutron sphere.
These Figures also verify that for high enough stiff DM EOS for the DS2, the radius of DS2 is larger than the visible radius of star.
Our results confirm that the surface
of the neutron sphere can lie between the surfaces of DS1 and DS2.
For example considering the DM EOS $p_0=0.6\times10^{35}dyncm^{-2}$, for the star
with $M=0.54\ M_{\odot}$ we have obtained $R_N=8.24\ km$, $R_{D1}=8.01\ km$, and $R_{D2}=11.99\ km$.
Besides, the total mass and the radius of DS2 assure the Buchdahl condition, i.e. $M < 4R_{D2}/9$,
and the stability is approved.

\subsection{Neutron sphere mass versus the visible radius}
\begin{figure*}[t]
\centering{%
{  \includegraphics[width=1.0\textwidth]{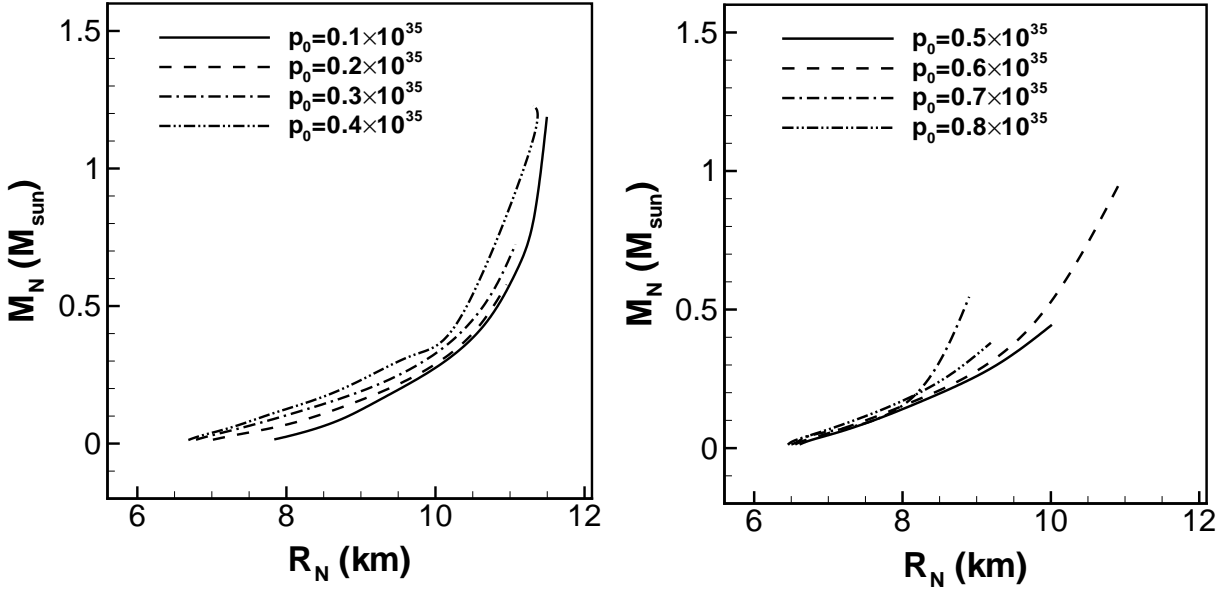}\qquad}}
\caption{Mass of neutron sphere, $M_N$, as a function of neutron sphere radius, $R_N$,
with the DM EOS $p_0=0.4\times10^{35}dyncm^{-2}$ for the DS1 and different DM EOSs for the DS2.}
\label{fig9}
\end{figure*}

Fig. \ref{fig9} presents the contribution of neutron portion in the mass of star, $M_N$,
versus the radius of neutron sphere in the case of DM EOS $p_0=0.4\times10^{35}dyncm^{-2}$ for the DS1 and different DM EOSs for the DS2. For the stars with
larger neutron sphere, the mass of neutron sphere is higher. It is clear that
with a mass for neutron sphere, the neutron sphere radius decreases by increasing
the stiffness of DM in DS2. Fig. \ref{fig9} indicates that with the DM EOS $p_0\lesssim0.4\times10^{35}dyncm^{-2}$, the contribution of neutron sphere in the star mass is more significant compared to stiff EOSs.
For example in the case with $p_0=0.4\times10^{35}dyncm^{-2}$, for the star with total mass $M=1.32M_{\odot}$ and
visible radius $R_N=11.31\ km$, the mass of neutron sphere has been obtained $M_N=1.23M_{\odot}$.
Figs. \ref{fig3} and \ref{fig9} verify that for $p_0\lesssim0.4\times10^{35}dyncm^{-2}$ the behavior of the neutron sphere mass and total mass versus the size of neutron sphere are similar.
The mass of neutron sphere and neutron sphere radius convince the Buchdahl condition, i.e. $M_N < 4R_N/9$,
and the stability is confirmed.

\subsection{Mass and radius of dark matter spheres}
\begin{figure*}[t]
\centering{%
{   \includegraphics[width=1.0\textwidth]{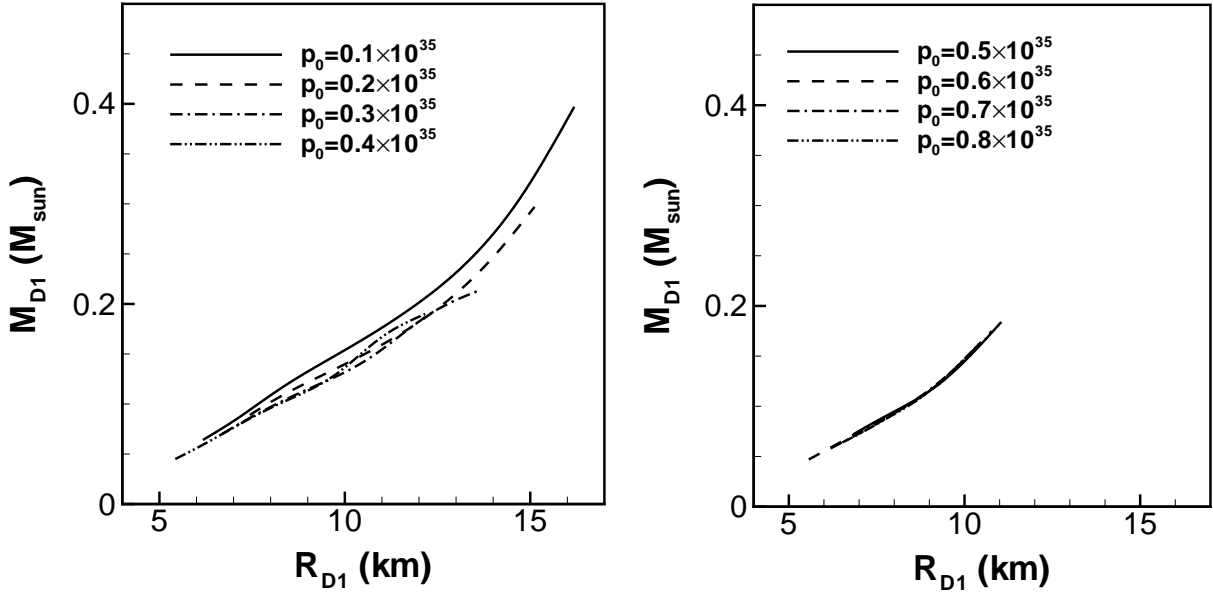}\qquad}}
\caption{Mass of DS1, $M_{D1}$, versus its radius, $R_{D1}$,
with the DM EOS $p_0=0.4\times10^{35}dyncm^{-2}$ for the DS1 and different DM EOSs for the DS2.}
\label{fig11}
\end{figure*}

Fig. \ref{fig11} gives the mass of DS1 versus its radius for the DM EOS $p_0=0.4\times10^{35}dyncm^{-2}$ for the DS1 and different DM EOSs for the DS2.
For the stars with large DS1, the contribution of this sector
in the total mass is more considerable.
We can see from Fig. \ref{fig11} that with the DM EOS $p_0\lesssim0.4\times10^{35}dyncm^{-2}$,
the DS1 can be more massive. In addition, the range of DS1 mass
is $0.05M_{\odot} \lesssim M_{D1} \lesssim 0.40M_{\odot}$ which is less than the range $0.01M_{\odot} \lesssim M_N \lesssim 1.23M_{\odot}$ related to the neutron sphere (see Fig. \ref{fig9}). Fig. \ref{fig11} indicates
that for $p_0>0.4\times10^{35}dyncm^{-2}$, the $M_{D1}-R_{D1}$ relation is not significantly affected by the DM EOS. With the softer DM in the DS2, the radius of DS1 can be higher, i.e. $R_{D1}>16\ km$. Figs. \ref{fig9} and \ref{fig11} explain that the radius of DS1 lies between $5\ km \lesssim R_{D1} \lesssim 16\ km$ which is a wider range
in comparison with $7\ km\lesssim R_N \lesssim 12\ km$ for the neutron sphere.
This phenomenon, along with the one related to the
mass of spheres results in the less accumulation of DM in comparison
with the visible matter. The mass of DS1 and its radius persuade the Buchdahl condition, i.e. $M_{D1} < 4R_{D1}/9$,
and the stability is verified.

\begin{figure*}[t]
\centering{%
{
  \includegraphics[width=1.0\textwidth]{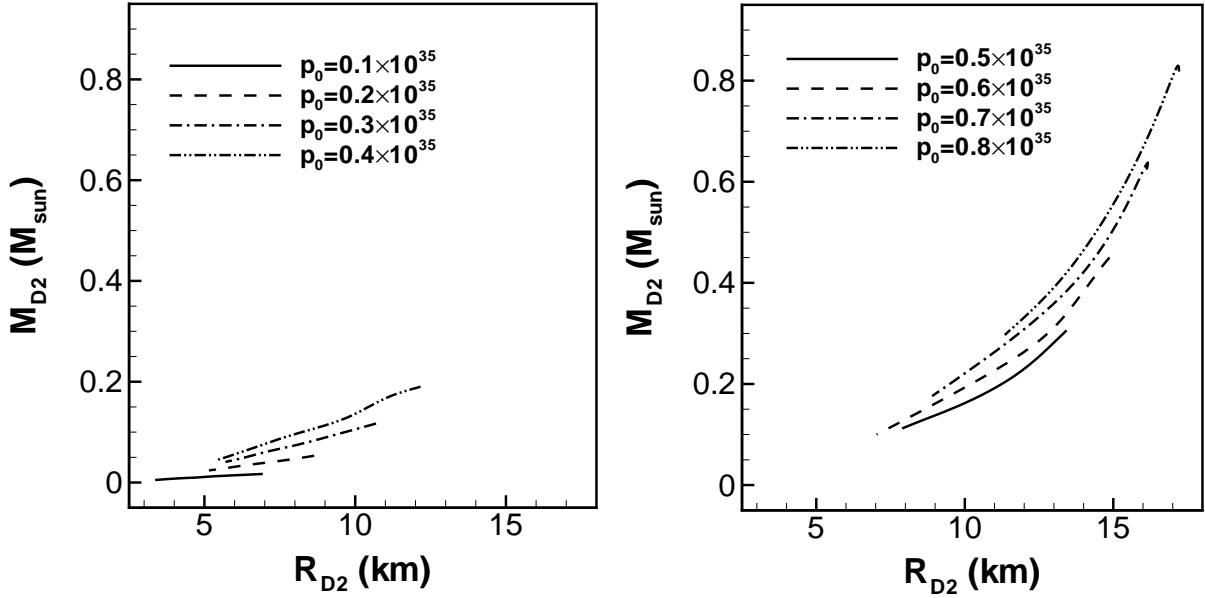}\qquad}}
\caption{Mass of DS2, $M_{D2}$, versus its radius, $R_{D2}$,
with the DM EOS $p_0=0.4\times10^{35}dyncm^{-2}$ for the DS1 and different DM EOSs for the DS2.}
\label{fig13}
\end{figure*}

In Fig. \ref{fig13}, we have plotted the mass of DS2 versus its radius
for the stars with the DM EOS $p_0=0.4\times10^{35}dyncm^{-2}$ for the DS1 and different DM EOSs for the DS2.
The larger the $R_{D2}$, the higher the $M_{D2}$, similar to the $M_{D1}-R_{D1}$ relation. However, unlike the DS1, for the stiffer DM EOSs, the DS2 has larger size and higher mass compared to the cases $p_0<0.4\times10^{35}dyncm^{-2}$.
For example, with the DM EOS $p_0=0.8\times10^{35}dyncm^{-2}$, for the star with total mass $M=1.01M_{\odot}$ and
visible radius $R_N=6.55\ km$, the mass and radius of DS2 are $M_{D2}=0.82M_{\odot}$ and
$R_{D2}=17.21\ km$. We can conclude that for $p_0>0.4\times10^{35}dyncm^{-2}$, the contribution of DS2 in the total mass is more significant than the DS1. With the stiffer DM in the DS2, the mass of DS2 increases. Fig. \ref{fig13} also verifies that the slope of $M_{D2}-R_{D2}$ relation is higher with the stiffer DM EOSs.
Comparing Figs. \ref{fig11} and \ref{fig13} shows that the influence of DM EOS of the DS2
on $M_D-R_D$ relation is more considerable for the DS2. We can see from Figs. \ref{fig9} and \ref{fig13} that for $p_0\lesssim0.6\times10^{35}dyncm^{-2}$, the contribution of DS2 in the total mass is smaller than the neutron sphere. But with $p_0>0.6\times10^{35}dyncm^{-2}$, the DS2 mass can be higher than the neutron sphere mass.
The radius of DS2
lies within $3\ km\lesssim R_{D2}\lesssim 17\ km$, a wider range compared to the neutron sphere. This shows the less accumulation of DM in DS2.
However, the almost similar range of radius for two dark spheres
but higher values of the mass of DS2 indicates the
more possible condition for the DM accumulation in the DS2
compared to DS1. The mass of DS2 and its radius provide the Buchdahl condition, i.e. $M_{D2} < 4R_{D2}/9$,
and the stability is approved.

\subsection{Double dark-matter admixed neutron stars with different neutron and dark matter contributions}

\begin{figure*}[t]
\centering{%
{
  \includegraphics[width=1.0\textwidth]{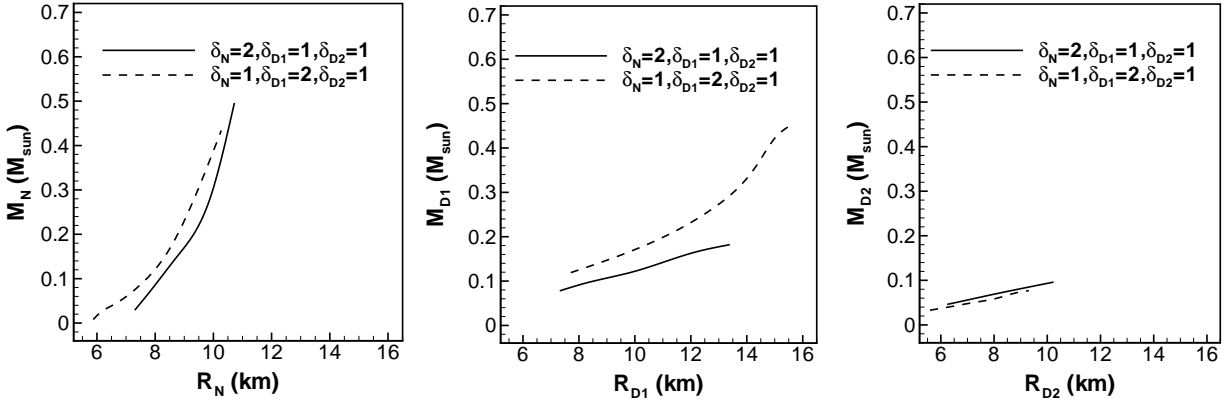}\qquad}}
\caption{Mass-Radius relations for three spheres of neutron stars with different contributions of neutron matter and DM in DS1. The DM EOSs are $p_0=0.4\times10^{35}dyncm^{-2}$ for the DS1 and $p_0=0.3\times10^{35}dyncm^{-2}$ for the DS2.}
\label{fig15}
\end{figure*}

\begin{figure*}[t]
\centering{%
{
   \includegraphics[width=1.0\textwidth]{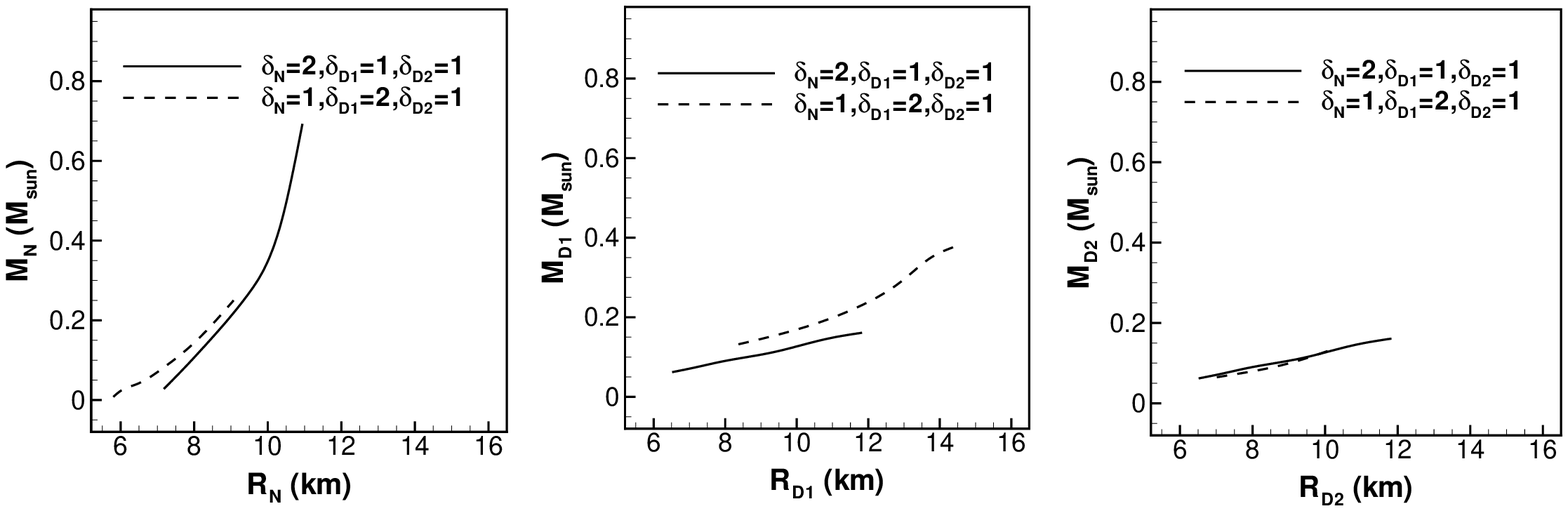}\qquad}}
\caption{Same as Fig. \ref{fig15} but for the DM EOSs $p_0=0.4\times10^{35}dyncm^{-2}$ for the DS1 and also $p_0=0.4\times10^{35}dyncm^{-2}$ for the DS2.}
\label{fig16}
\end{figure*}
\begin{figure*}[t]
\centering{%
{
  \includegraphics[width=1.0\textwidth]{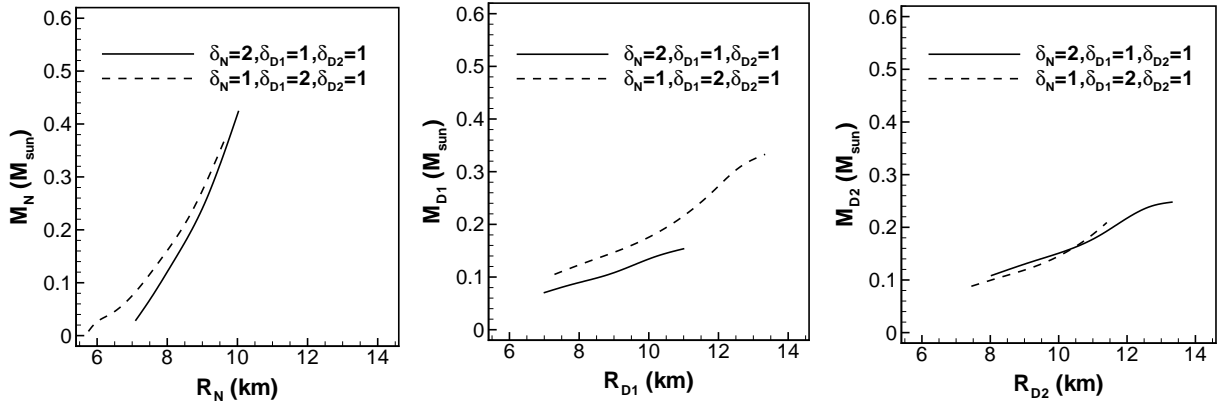}\qquad}}
\caption{Same as Fig. \ref{fig15} but for the DM EOSs $p_0=0.4\times10^{35}dyncm^{-2}$ for the DS1 and $p_0=0.5\times10^{35}dyncm^{-2}$ for the DS2.}
\label{fig17}
\end{figure*}

As mentioned above, the DM in the neutron star can exist from the first steps of star formation or
it can be captured by the neutron star. Therefore, the contribution of neutron matter and DM in DS1 may be different from each other. This can be presented by different central pressures of neutron matter and DM in DS1. We define the parameter $\delta_i=\frac{p_{i}(r=0)}{p_c}$ for the sector $i$, in which $p_c$
denotes the central pressure which has been equal for three sectors heretofore, and $i=N$,$D1$, and $D2$, respectively. We consider the stars with different sets of  $\delta_i$, i.e. $\delta_N=2,\delta_{D1}=1,\delta_{D2}=1$ and $\delta_N=1,\delta_{D1}=2,\delta_{D2}=1$. The first set shows the stars which are primarily made with the visible matter and then captured the DM, first-visible stars. The second case indicates the stars with the initial interstellar cloud with the DM as its main content and then captured the visible matter in its evolution, first-dark stars. We assume that all these stars indicated by these two sets are doubly affected by the DM in a similar condition with $\delta_{D2}=1$.
Figs. \ref{fig15}-\ref{fig17} present the mass versus the radius for three spheres
of neutron stars with different sets of $\delta_i$ for DM EOS $p_0=0.4\times10^{35}dyncm^{-2}$ in the DS1 and different DM EOSs for the DS2.
For all DM EOSs for the DS2, the neutron sphere of first-dark stars is smaller than this
sphere in first-visible stars. The mass of neutron matter sphere takes higher values for the first-visible stars, especially for the case $p_0=0.4\times10^{35}dyncm^{-2}$ in the DS2. Moreover, as we expected, the DS1 mass is higher for the first-dark stars. However, for the most stars, the DS2 mass is slightly lower for the first-dark stars. This indicates that the first-visible stars can capture more DM in
the DS2 compared to the first-dark stars. The radius of the DS1 in the first-dark stars can be larger than this radius in the first-visible stars. However, the range of DS1 radius depends on the DM EOSs for the DS2. By increasing the stiffening of the DM EOS for the DS2, the DS1 radius for both first-visible and first-dark stars can take smaller values. Noting the results for the radius of DS2, we conclude that in the first-visible stars, the DS2 radius can be higher than this radius in the first-dark stars. Besides, with the stiffer DM in the DS2, the DS2 radius for both first-visible and first-dark stars gets the higher values. We can see from Figs. \ref{fig15}-\ref{fig17} that for all DM EOSs and different contributions of
neutron matter and DM in DS1, the Buchdahl conditions for all spheres, i.e. $M_N < 4R_N/9$ and
$M_{D1} < 4R_{D1}/9$ and $M_{D2} < 4R_{D2}/9$, are satisfied and these spheres are stable.

\subsection{Gravitational redshift versus the total mass}

The gravitational redshift at the surface of neutron star, the criterion for the star compactness, is
given by
\begin{eqnarray}
Z_s=[1-2(\frac{M}{R})]^{-1/2}-1,
\end{eqnarray}
where R is the visible radius of the neutron star.
\begin{figure*}[t]
\centering{%
{
  \includegraphics[width=0.9\textwidth]{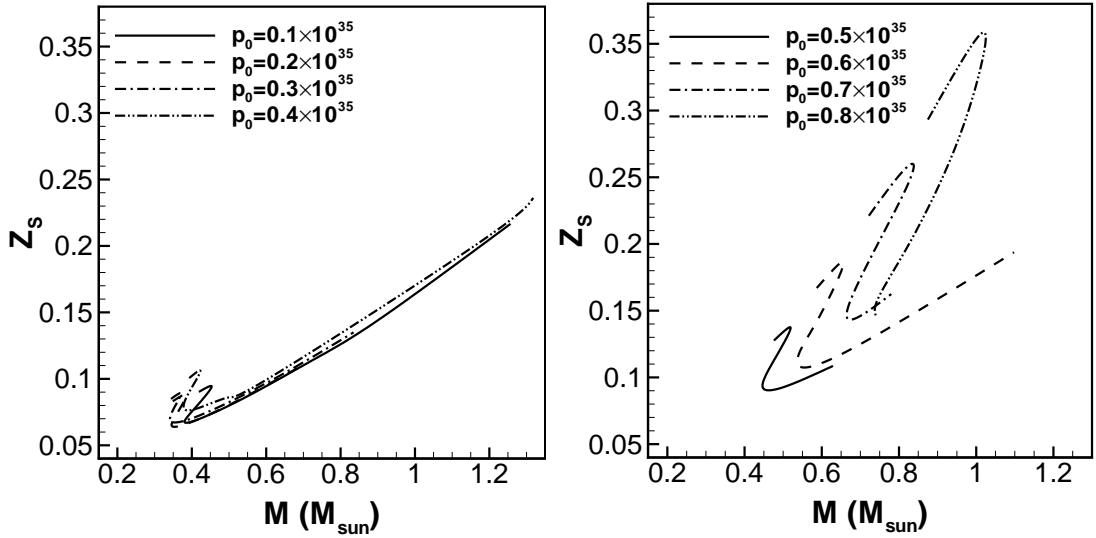}\qquad}}
\caption{Gravitational redshift at the surface of neutron star, $Z_S$, considering the
equal central pressures of neutron and DM in two spheres and
the DM EOS $p_0=0.4\times10^{35}dyncm^{-2}$ for the DS1 and different DM EOSs for the DS2.}
\label{fig21}
\end{figure*}
We have shown the gravitational redshift at the surface of neutron stars considering the
equal central pressures of neutron matter and DM with
the DM EOS $p_0=0.4\times10^{35}dyncm^{-2}$ for the DS1 and different DM EOSs for the DS2 in Fig. \ref{fig21}. With the stiffer DM in DS2, the value of the gravitational redshift is higher. With the DM EOS $p_0=0.8\times10^{35}dyncm^{-2}$, $Z_S$ can be about $0.35$.
The effects of the DM EOSs of the DS2 on the gravitational redshift is more significant with $p_0>0.4\times10^{35}dyncm^{-2}$. Since the gravitational redshift is obtained using the observational data,
it is possible to use our results to study the double dark-matter admixed neutron stars.

\section{Summary and Concluding Remarks}
The properties of neutron stars affected by the dark matter via two
different processes (primitive and secondary DM) have been considered. The general relativistic formalism and
the dark matter equation of state from pseudo-isothermal model have been
employed.
The mass-radius relation of double dark-matter admixed neutron star
affects by the degree of the stiffness of DM they absorb.
The stars with the stiffer DM can be smaller and more massive.
With softer secondary DM, the radius of primitive dark matter halo increases
by decreasing the star mass. However, for the stiff secondary DM,
this radius increases when the star mass grows.
For the massive stars with softer secondary DM, the primitive DM is surrounded by the
neutron matter sphere. But, in the stars with lower masses and softer secondary DM, this DM has a size larger
than the neutron matter sphere.
This is while with the stiff secondary DM, the condition is vice versa.
The contribution of neutron, primitive DM, and secondary DM in the star mass depends on the
stiffness of secondary DM.
We have found that the neutron matter sphere of first-dark stars
 is smaller compared to first-visible stars. In addition. the mass of
neutron matter sphere takes higher values for the first-visible stars.
It has been shown that the first-visible stars can capture more DM in the secondary process
compared to the first-dark stars.
Moreover, in the first-visible stars, the secondary DM
radius can be higher than this radius in the first-dark stars. Our calculations show that for the double dark-matter admixed model of neutron star with different or the same contributions of neutron matter and DM in DS1, all the spheres,
i.e. neutron sphere and DS1 and DS2, satisfy the Buchdahl condition and all the spheres are stable.
Finally, we have found that
with more stiff secondary DM, the
value of the gravitational redshift increases.

\section*{Acknowledgements}
The author wishes to thank the Shiraz University Research Council.

\end{document}